\documentclass [onecolumn]{aa}
\usepackage{graphicx}

\begin{document}

    \title{A method of open cluster membership determination}

    \author{G. Javakhishvili, V. Kukhianidze, M. Todua \and R. Inasaridze }

    \institute{Abastumani Astrophysical Observatory, Georgian Academy of Sciences\\
                \email{todua@hotmail.com}}


    \begin{abstract}
    {
    A new method for the determination of open cluster membership based
    on a cumulative effect is proposed. In the field of a plate the
    relative $x$ and $y$ coordinate positions of each star with respect
    to all the other stars are added. The procedure is carried
    out for two epochs $t_1$ and $t_2$ separately, then one sum is
    subtracted from another. For a field star the differences in its relative coordinate
    positions of two epochs will be accumulated. For a cluster
    star, on the contrary, the changes in relative positions of cluster
    members at $t_1$ and $t_2$ will be very small. On the histogram of sums the
    cluster stars will gather to the left
    of the diagram, while the field stars will form a tail to the right.
    The procedure allows us to efficiently discriminate one group from
    another. The greater the distance between $t_1$ and $t_2$ and the more
    cluster stars present, the greater is the effect. The
    accumulation method does not require reference stars,
    determination of centroids and modelling the distribution of
    field stars, necessary in traditional methods. By the proposed method 240
    open clusters have been processed, including stars up to $m<13$. The membership
    probabilities have been calculated and compared to those obtained by the most
    commonly used Vasilevskis-Sanders method. The similarity of the results acquired the
    two different approaches is satisfactory for the majority
    of clusters.

    }
    \end{abstract}

    \authorrunning{Javakhishvili et al.}
    \titlerunning{Open claster membership}

    \maketitle

\section{Introduction}
    Various methods based on the analysis of positions, proper motions, radial
    velocities, magnitudes and their combinations have been
    proposed to determine the members of open clusters.

    The first mathematically rigorous procedure for determination of
    open cluster membership was developed by Sanders (\cite{Sanders})
    with a statistical analysis of proper motions. It is also
    the most widely used method. Sanders's approach is based on the
    model of overlapping distributions of field and cluster stars in
    the neighborhood and within the region of visible grouping of
    stars, introduced by Vasilevskis et al. (\cite{Vasilevskis}).

    Vasilevskis's model implies that proper motion dispersion of cluster
    members is caused by observational and measurement errors
    assumed to be normally distributed. Thus the distribution of cluster stars is represented
    by a bivariate normal frequency function. The dispersion of field
    stars is due to not only the errors referred to above, but also to peculiar motion
    and differential galactic rotation. Therefore the field star distribution is
    not expected to be random, but rather to have a preferential direction
    and not normal distribution. However, in a first approximation a bivariate normal
    ellipsoidal distribution function was assumed for field stars, the major axis of the
    ellipse being parallel to the galactic plane.

    Thus, Sanders's equations contain 8 unknown parameters to be determined:
    the number of cluster members, x and y components of two centroids, one circular and
    two elliptical dispersions. This system is solved by a maximum
    likelihood method (Sanders \cite{Sanders}). The probability of stars being
    cluster members is calculated by frequency functions with determined
    parameters.

    Slovak (\cite{Slovak}) tested the Vasilevskis-Sanders method by
    modelling the proper motion distribution in the surroundings of a
    cluster. He proved the uniqueness and convergence of solutions
    of the system, provided that errors are represented by a
    Gaussian distribution and open clusters have no significant internal
    motion. But if there is noticeable motion within a cluster or it
    rotates then the above method fails.

    Cabrera-Ca\~{n}o and Alfaro (\cite{Cabrera-Cano Alfaro1})
    improved the numerical techniques for obtaining the above parameters. McNamara
    and Schneeberger (\cite{McNamara Schneeberger}) showed that the final
    probabilities could be influenced by various weight groups.
    Zhao and He (\cite{Zhao He}) provided a method for treating
    data with different accuracies.

    However all these improvements did not treat the problem of star distribution
    model on which the membership probabilities are based. Even if the
    hypothesis of a normal distribution of field stars is realistic for some clusters, the
    centroids of field and cluster stars sometimes are too close to be well
    discriminated. The parametric model also fails when the cluster member-to-field
    star ratio is small. The Vasilevskis-Sanders method does not
    work in the case of significant internal motion in a cluster or its rotation.

    To overcome some of the problems arising from the parametric
    Vasilevskis-Sanders method, especially the star distribution modelling,
    Cabrera-Ca\~{n}o and Alfaro (\cite{Cabrera-Cano Alfaro2})
    developed a more general, non-parametric method of membership
    determination. Here no assumptions were made
    about the nature of cluster and field star distributions and allowed
    for the use of photometric data. Each cluster needed careful individual study.

    So, the actual distribution of cluster
    and field stars, especially the latter, may not be fitted by a Vaselevskis-Sanders model;
    the centroids for the two groups may be too close to be distinguished; the reliability
    of the method depends on the cluster-to-field star ratio; in the case of significant internal
    motions or rotation of a cluster the traditional method fails.

    In the next section we introduce a new method for the discrimination of cluster stars
    from surrounding field stars based on a cumulative effect using positions and proper
    motions. We do not try to overcome all the difficulties of traditional methods, though we
    offer another approach that enlarges the statistical distance between the two
    populations - cluster members and field stars - by revealing the group of stars with
    the least relative velocities. The advantage of this method is
    that no assumption is made about the distribution of field stars and determination of
    centroids is avoided. However, in order to determine probabilities we still have to
    assume a normal bivariate distribution for clusters. Another advantage of the method
    is that reference stars are not necessary: the discrimination of cluster members is most
    effective by rectangular coordinates. These features of the method allow us to increase
    the statistical distance between the two populations. The most
    noticeable advantage of the accumulation method is its ability to reveal dynamic
    structures within the clusters if there are any.

\section{The method}
We consider a cluster as a gravitationally bound system with a
bulk motion relative to surrounding galactic field stars. Field
stars are characterized by more or less random velocities.
Therefore the cluster members must show regular motion, while
the field stars move realtively irregularly.

We consider rectangular coordinates of stars in the field of a
plate (or CCD array), measured in two epochs $t_1$ and $t_2$. The
relative $x$ and $y$ positions of each star with respect to all the
other stars in the area are added. If the two epochs are well
separated, then for a field star the differences of the relative
coordinate positions in the two epochs will be accumulated. For
cluster stars, the changes of relative positions of physical members
at $t_1$ and $t_2$ must be very small and their sum will be minimal.
The more cluster stars, the bigger is the effect. If
one draws a histogram of differences, the stars with minimal
relative velocities (which we consider more likely to be cluster
members) will gather to the left of the diagram, while the field
stars will form a tail to the right. Thus, the procedure allows us to
efficiently discriminate one group from another, avoiding the use of
reference stars.

For each star the following quantities are calculated:

\begin{equation}
    S_{x_i}^\prime = \sum _{j=1}^N \vert {x_i^\prime - x_j^\prime}
    \vert,
\\
    S_{y_i}^\prime = \sum _{j=1}^N \vert y_i^\prime - y_j^\prime
    \vert,
\end{equation}
\begin{equation}
    S_{x_i}^{\prime \prime} = \sum _{j=1}^N \vert x_i^{\prime
    \prime} - x_j^{\prime \prime} \vert,
\\
    S_{y_i}^{\prime \prime} = \sum _{j=1}^N \vert y_i^{\prime \prime} -
    y_j^{\prime \prime} \vert,
\end{equation}
where $N$ is a total number of stars (on the plate or CCD array).
The primed quantities belong to the first epoch, the double primed
to the second one.

Then the differences of $S_{x_i}^\prime$ and $S_{x_i}^{\prime
\prime}$ are calculated:

\begin{equation}
    \delta S_{x_i} = S_{x_i}^\prime - S_{x_i}^{\prime \prime},
\\
    \delta S_{y_i} = S_{y_i}^\prime - S_{y_i}^{\prime \prime},
\\
    i=1, ..., N
\end{equation}

The next step is to draw the histograms for $\delta S_{x_i}$ and
$\delta S_{y_i}$ separately. According to our assumption, cluster
stars will concentrate within the first bin with minimal $(\delta
S_{x_i}, \delta S_{y_i})$.

In order to define the probabilities of belonging stars to a cluster,
the first one or two most populated bins are chosen. Then the whole
procedure is repeated for the stars from the selected bins using the
formulae (1) and (2), but dropping the moduli this time, that is:

\begin{equation}
    \tilde S_{x_i}^\prime = \sum _{j=1}^N (x_i^\prime - x_j^\prime),
\\
    \tilde S_{y_i}^\prime = \sum _{j=1}^N (y_i^\prime - y_j^\prime),
\end{equation}
\begin{equation}
    \tilde S_{x_i}^{\prime \prime} = \sum _{j=1}^N (x_i^{\prime \prime} - x_j^{\prime
    \prime}),
\\
    \tilde S_{y_i}^{\prime \prime} = \sum _{j=1}^N (y_i^{\prime \prime} - y_j^{\prime
    \prime}),
\end{equation}

Again, the histograms are drawn. To determine the probabilities of stars
being physical members, we use the normal ellipsoidal distribution function
as the best fit for a cluster in the first approximation. So, the obtained diagrams are
fitted with Gaussian curve by a least square method:

\begin{equation}
    f(x) =\frac{A_x}{w_x \sqrt {\pi / 2} } e^{{-2 ({\frac
    {x-x_0}{\sigma_x}})^2}},
\\
    f(y) =\frac{A_y}{w_y \sqrt {\pi / 2} } e^{{-2 ({\frac
    {y-y_0}{\sigma_y}})^2}},
\end{equation}
where $A$ is an area of the normal distribution curve, $\sigma$ is a
double dispersion.

The probabilities by $x$ and $y$ coordinates for each star are
calculated as:

\begin{equation}
    p_x = e^{{-2 ({\frac {x-x_0}{\sigma_x}})^2}},
\\
    p_y = e^{{-2 ({\frac {y-y_0}{\sigma_y}})^2}},
\end{equation}

The resulting probability is:
\begin{equation}
    p = p_x * p_y
\end{equation}

If there is no "raw material" (i.e. rectangular coordinates)
available, but there are proper motions instead, then (1), (2),
(4) and (5) are modified as follows:

\begin{equation}
    \delta \mu_{\alpha_i} = \sum _{j=1}^N \vert \mu_{\alpha_i} - \mu_{\alpha_j}
    \vert,
\\
    \delta \mu_{\delta_i} = \sum _{j=1}^N \vert \mu_{\delta_i} - \mu_{\delta_j} \vert
\end{equation}

\begin{equation}
    \tilde \delta \mu_{\alpha_i} = \sum _{j=1}^N  (\mu_{\alpha_i} -
    \mu_{\alpha_j}),
\\
    \tilde \delta \mu_{\delta_i} = \sum _{j=1}^N  (\mu_{\delta_i} -
    \mu_{\delta_j})
\end{equation}

The rest of the procedure remains the same. In this article we apply this modified
version of the method to the clusters from Dias's (\cite{Dias1}, \cite{Dias2})
catalogues and compare the results.

As will be shown later, the accumulation method can reveal various dynamic
groups in a cluster, if there are any.

\section{Results}

\begin{figure}
\centering
\includegraphics[width=10cm]{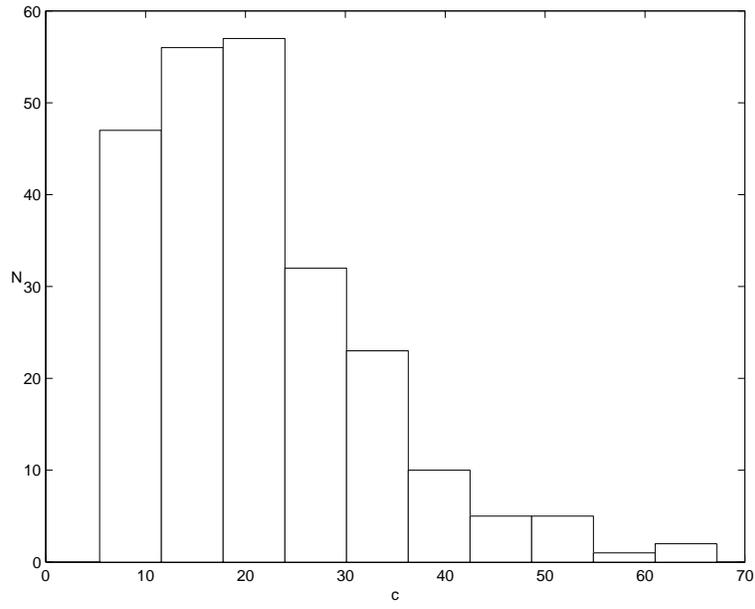}
    \caption{Histogram for $C$ - the degree of similarity of the results obtained by Dias and the accumulation method (11).}
    \label{FigHistC}
\end{figure}

The accumulation method was applied to 240 open clusters of
the Dias catalogues, including the
stars up to $m<13$. The clusters in the Dias list were processed by
a Vasilevskis-Sanders approach. In order to compare our results, we
introduce a degree of similarity for each cluster as:

\begin{equation}
    C = \frac {\sum_{i=1}^{n} \vert P_{d_i} - P_{o_i} \vert}{n}
\end{equation}
where $P_d$ are probabilities by Dias,  $P_o$ - ours, $n$ - number
of stars in a cluster. For $C$ a histogram was drawn (see
Fig.~\ref{FigHistC}). The smaller the $C$ the closer are the results
obtained by the two different methods.

As one can see from the histogram, the similarity of probabilities
is quite satisfactory for the great majority of
clusters: 162 out of 240 have $C<25$, 65 clusters have $C$
in the range between 25 and 45 and 13 of them $45<C<70$. The latter group
have the most serious discrepancies - almost an anticorrelation.

\begin{figure}
\centering
\includegraphics[width=\textwidth]{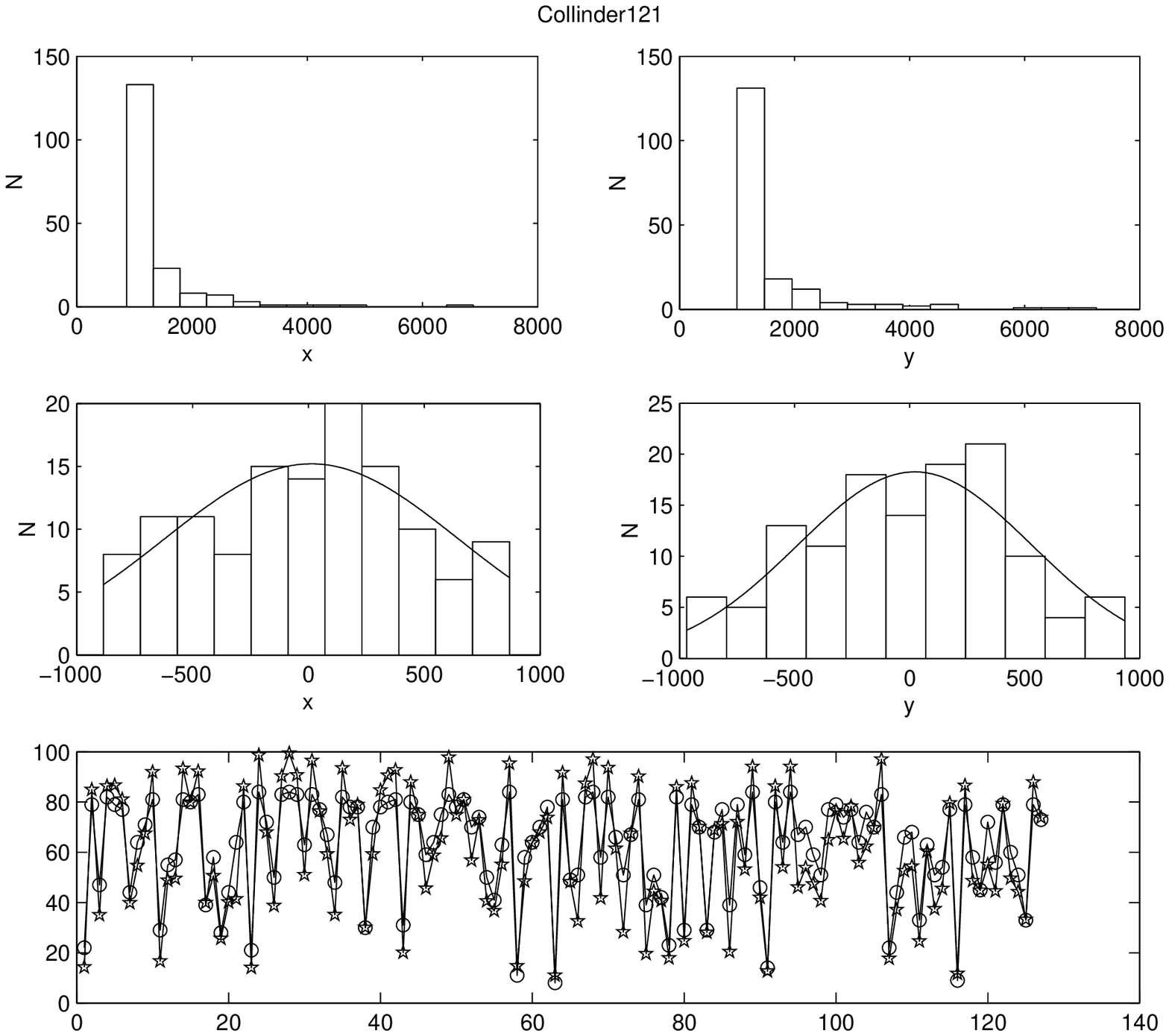}
    \caption{Collinder 121: the first row - the
histograms for $\vert \delta \mu_\alpha
\vert$ and $\vert \delta \mu_\beta \vert$
(formulae (9)); the second row - histograms for $\delta \mu_\alpha$
and $\delta \mu_\beta$ (formulae (10)); the
third row - comparison of individual probabilities by Dias
(circles) and by the accumulation method (asterisks)}
    \label{Figcollinder121}
\end{figure}

\begin{figure}
\centering
\includegraphics[width=\textwidth]{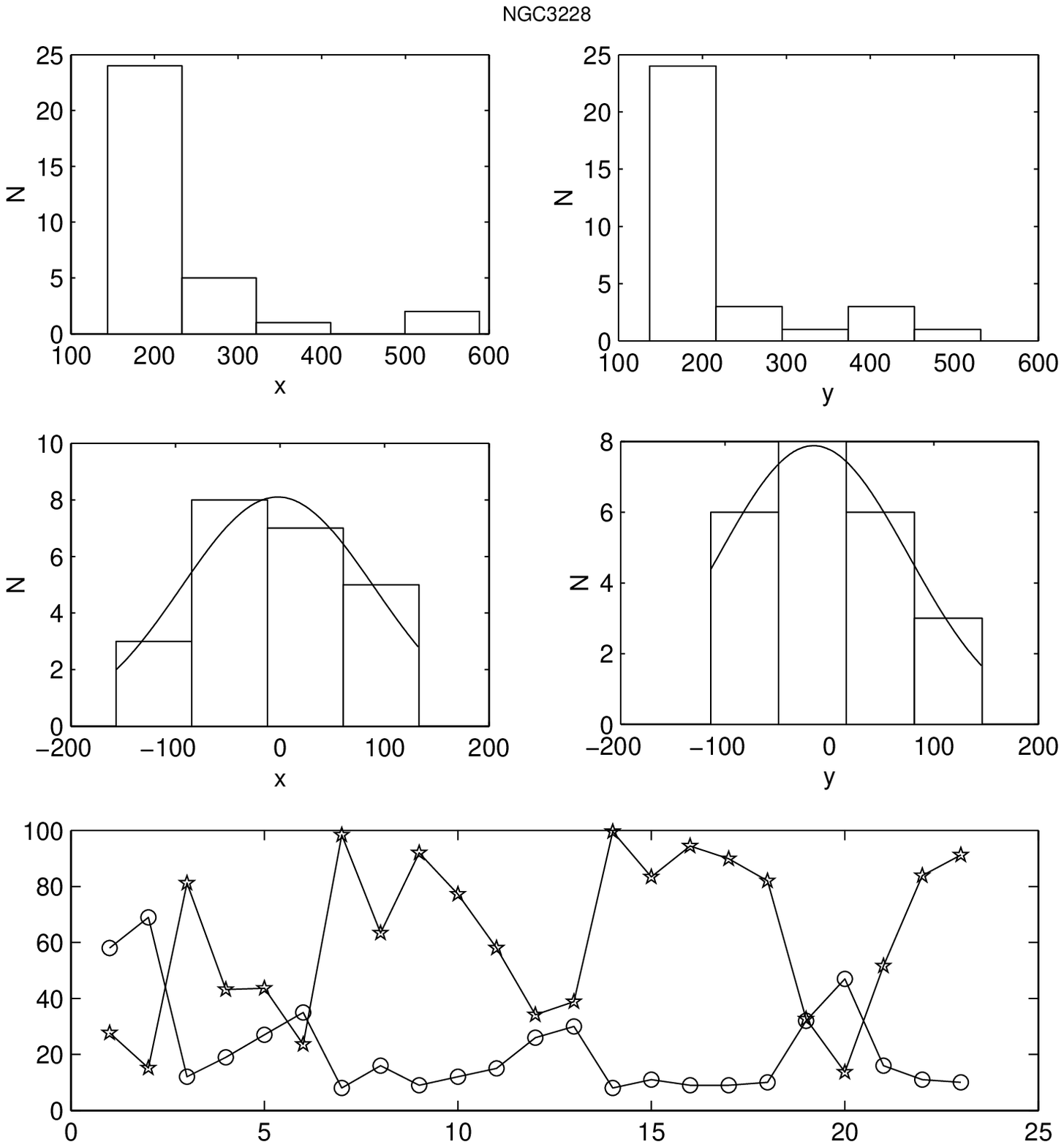}
    \caption{NGC3228: the first row - the
histograms for $\vert \delta \mu_\alpha
\vert$ and $\vert \delta \mu_\beta \vert$
(formulae (9)); the second row - histograms for $\delta \mu_\alpha$
and $\delta \mu_\beta$ (formulae (10)); the
third row - comparison of individual probabilities by Dias
(circles) and by the accumulation method (asterisks)}
    \label{FigNGC3228}
\end{figure}

We choose two cases to show: Collinder 121
(Fig.~\ref{Figcollinder121}), with $C=7$, total number of stars $n=179$ - good
agreement, and NGC 3228 (Fig.\ref{FigNGC3228}), with $C=50$,
$n=32$ - disagreement.

In Fig.~\ref{Figcollinder121} and Fig.\ref{FigNGC3228}, the
histograms in the upper row are for $\vert \delta \mu_\alpha
\vert$ and $\vert \delta \mu_\beta \vert$ according to the
formulae (9), those in the middle row - for $\delta \mu_\alpha$
and $\delta \mu_\beta$ by the formulae (10). In the figure on the
third row the individual probabilities by Dias
(denoted by circles) and by the accumulation method (asterisks) are
presented.

\begin{figure}
\centering
\includegraphics[width=\textwidth]{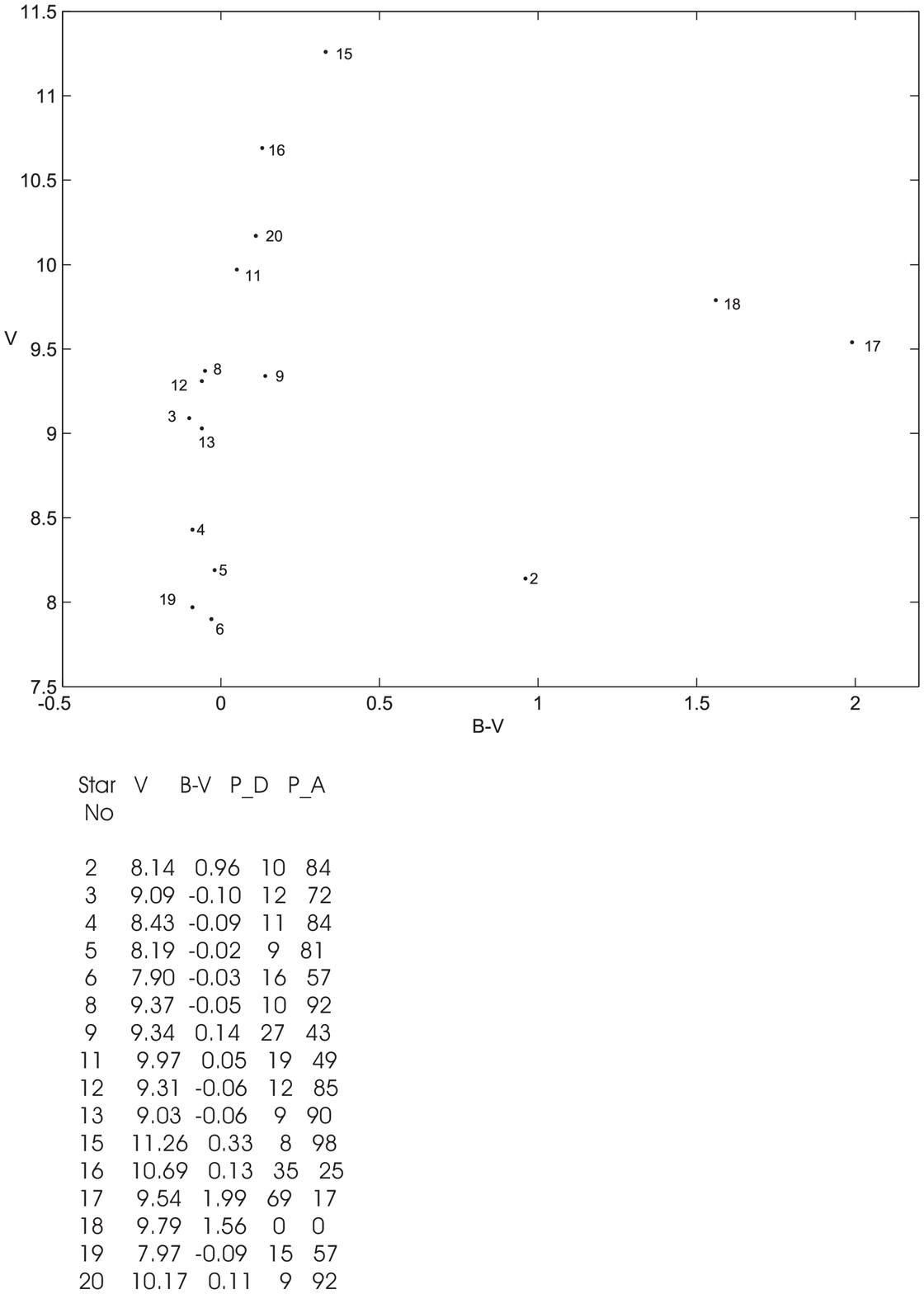}
    \caption{The color-magnitude diagram for NGC3228}
    \label{FigCMNGC3228}
\end{figure}

The probabilities for the cluster Collinder 121 (Fig.~\ref{Figcollinder121}) reveal close
similarity, as for 2/3 of clusters, while those of the cluster NGC 3228 (Fig.\ref{FigNGC3228})
differ considerably from Dias's results. In the latter case the
anticorrelation of probabilities is striking: the stars we claim
as cluster members are field stars by Dias and vice versa. In this case
(number of stars 32) the probabilities are on average smaller than ours.
Our probabilities are somewhat more pronounced: cluster stars have
high probabilities, while those by Dias are low.
According to Dias, it may not be considered as a cluster at all, since only two stars have
membership probabilities greater then 50
(only 16 out of 32 have $V$ and $B-V$ quantities measured), as shown in
Fig.~\ref{FigCMNGC3228}. 13 stars out of 16 lie along one sequence confirming that this is
an open cluster with high probability. In the table below we give $V$ and $B-V$
magnitudes and probabilities by Dias ($P_D$) and by the accumulation method ($P_A$) for each star.
Star No 17 is claimed to be a cluster star by Dias, while we claim both stars Nos 17 and 18, which are
far from the main sequence of the diagram, to be field stars.

The reasons for the disagreements between our results in this case, as
well as in those few cases with poor accordance could be several:
the low cluster-to-field star ratio, or the presence of more then one dynamic structure
or large observational errors.

\begin{figure}
\centering
\includegraphics[width=16cm]{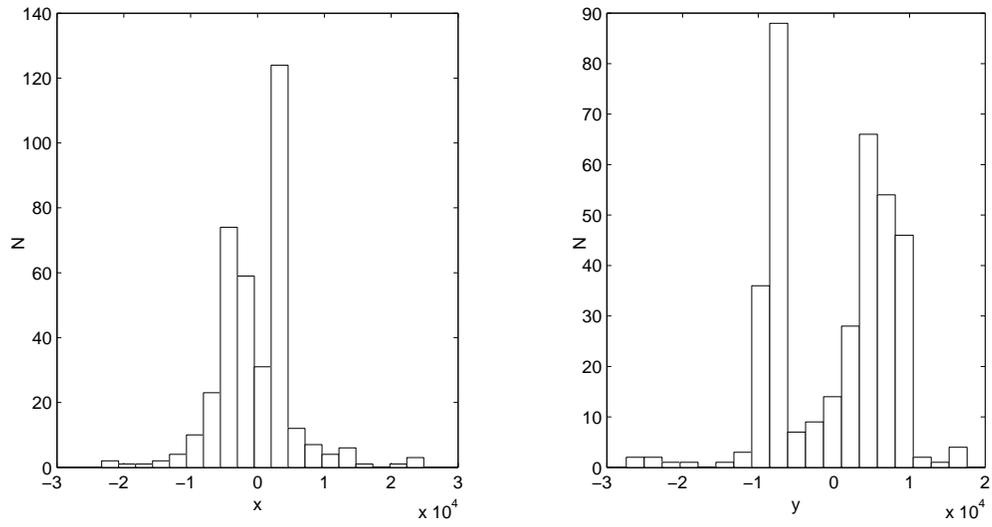}
    \caption{The Pleades (Mellotte22): histograms for $\delta \mu_{\alpha_i} $ (left) and
$\delta \mu_{\delta_i} $ (right).}
    \label{FigMel22}
\end{figure}

Pleades (Fig.\ref{FigMel22})is an example of dynamic structure or a rotation.
The histograms for $\delta mu_{\alpha_i}$ (upper diagram) and
$\delta mu_{\delta_i}$ (lower diagram) are shown, the quantities calculated using (9) and (10).
Two peaks on both diagrams are clearly seen, which reflects either two dynamic structures or
rotation of the cluster.

\section{Conclusion}

In the presented method of accumulation we assume that the cluster
members moving in the Galaxy as a whole have similar velocities, while
field stars show a wide range of velocities. The procedure - adding close
velocities while compensating for
disperced ones - should enhance the assembling of cluster stars, while
distributing more sparsely the field ones, due to the cumulative effect. Thus
this method effectively enlarges the statistical distance
between physical members and field stars, so that a membership as well as
a non-membership is more pronounced. This effect is better for larger
cluster-to-field star ratios. For field stars no particular distribution is
assumed, thus the centroids are not determined. However, when calculating
probabilities, for physical members a normal bivariate distribution function
is assumed. For rectangular coordinates no reference stars are needed in the
procedure. The most interesting feature of the accumulation mathod is its
ability to reveal more then one group of velocities, which has been shown with
the example of Pleades.

The modified accumulation method was applied to 240 clusters from Dias's
list. The probabilities calculated by the accumulation method showed
satisfactory agreement with those obtained by the Vasilevskis-Sanders method for
the majority of clusters. The poor agreements or disagreements can be ascribed
to low cluster-to-field star ratios, or multiple dynamic structures.

The accumulation method can be further expanded using other
variables like radial velocities and photometric quantities.

\clearpage

\begin{table}
\centering \caption {$V$ and $B-V$ magnitudes and probabilities for
individual stars of the cluster NGC3228.} \vspace{1 cm}
\begin{tabular}{|l|l|l|l|l|}
\hline
 Star No & V & $B_V$ & $P_D$ & $P_A$ \\
\hline
 2   & 8.14   &  0.96   & 10 & 84  \\
 3   & 9.09   &  -0.1   & 12 & 72  \\
 4   & 8.43   &  -0.09  & 11 & 84  \\
 5   & 8.19   &  -0.02   & 9 & 81  \\
 6   & 7.90   &  -0.03   & 16 & 57 \\
 8   & 9.37   &  -0.05   & 10 & 92 \\
 9   & 9.34   &  0.14   & 27 & 43  \\
 11   & 9.97   &  0.05   & 19 & 49 \\
 12   & 9.31   &  -0.06   & 12 & 85\\
 13   & 9.03   &  -0.06   & 9 & 90 \\
 15  & 11.26   &  0.33   & 8 & 98  \\
 16  & 10.96   &  0.13   & 35 & 25 \\
 17  & 9.54   &  1.99   & 69 & 17  \\
 18  & 9.79   &  1.56   & 0 & 0    \\
 19  & 7.97   &  -0.09   & 15 & 57 \\
 20  & 10.17   &  0.11   & 9 & 92  \\
\hline
\end{tabular}
\end{table}

\clearpage
\onecolumn


\newpage

\newpage

\newpage

\newpage

\end{document}